\documentclass[reprint,showpacs,aps,prl,amsmath,amssymb,floatfix]{revtex4-1}

\usepackage{graphicx,epstopdf}
\usepackage{bm}
\usepackage{color}
\usepackage{natbib}
\usepackage{CJK}

\allowdisplaybreaks[3]

\begin{document}
\begin{CJK*}{UTF8}{gbsn}

\title{Violating the no-signaling principle with classical inseparable beams\\in an optical parity-time symmetric system}

\author{Lida Zhang~(\CJKfamily{gbsn}张理达)}

\author{J\"{o}rg Evers}
\affiliation{Max-Planck-Institut f\"{u}r Kernphysik, Saupfercheckweg 1, D-69117 Heidelberg, Germany}

\date{\today}

\begin{abstract}
We show that the no-signaling principle can be violated with classical inseparable beams in the presence of a  parity-time (PT) symmetric subsystem. Thus, the problems associated to PT-symmetric quantum theories  recently discovered by  Lee et al. [Phys. Rev. Lett. {\bf 112}, 130404 (2014)] are not exclusive to quantum mechanics, but already exist in the classical case. The possibility to implement local optical PT-symmetric subsystems via light-matter interactions enables the experimental exploration of local PT symmetry and subtle quantum concepts via classical analogues.
\end{abstract}







\maketitle

\end{CJK*}

In canonical quantum mechanics, it is believed that Hamiltonians should be Hermitian, such that they have real eigenvalues and obey probability conservation. This axiom, however, is not derived from any fundamental physical principle. In the last decades, a new class of non-Hermitian Hamiltonians has been explored, which respect parity-time~(PT) symmetry and have a real eigenvalue spectrum~\cite{Bender1998PRL,Bender2007RPP}. This work is motivated by a potential significant impact on the foundations of quantum theory~\cite{Bender2002PRL,Bender2003AJP,Znojil2008PRD,Gong2013JPA,Mostafazadeh2010IJGMMP,Schmidt2013RSTA,Deffner2015PRL,Brody2016JPA} 
but also by fascinating applications in optical physics~\cite{Fleury2014PRL,Regensburger2012Nature,Lin2011PRL,
Feng2013NMat,Longhi2011JPA,Feng2014Science,Hodaei2014Science,Longhi2010PRA,Chong2011PRL,Sun2014PRL,Sukhorukov2010PRA,Chang2014Nphoton,Peng2014NPhys}.

Recently, Lee et al. have shown that local PT-symmetric quantum dynamics is incompatible with the no-signaling principle resulting from special relativity~\cite{Lee2014PRL}. Like in a standard Bell test with an entangled photon pair, their setup comprises a  source which emits two entangled photons to two remote observers Alice and Bob. Both observers can choose different local measurement settings $A$ and $B$. The no-signaling theorem then states that the local measurement statistics at one of the observers' sites should not be influenced by the choice of measurement of the other observer~\cite{Branciard2008NPhys}. Lee et al. showed that this principle can be violated, if a PT-symmetric Hamiltonian $\hat{H}$ locally acts on one of the two subsystems. In contrast, Hermitian $\hat{H}$ preserve the principle. To reach this result, two assumptions are made in the analysis: (1) The local quantum PT symmetric system is compatible with the conventional quantum theory; (2) The post-measurement process 
which is Hermitian is compatible with the PT symmetric evolution. Neglecting the possibility of superluminal signaling, one of the two assumptions must be violated. 
Lee et al. conclude that PT-symmetric quantum theory is not likely to be a fundamental theory of nature. 
Very recently,  the no-signaling principle was also experimentally tested using entangled photon pairs in an open quantum system which effectively simulates a PT-symmetric evolution~\cite{Tang2016NPhoton} by postselection.

\begin{figure}[b]
\includegraphics[width=8.0cm]{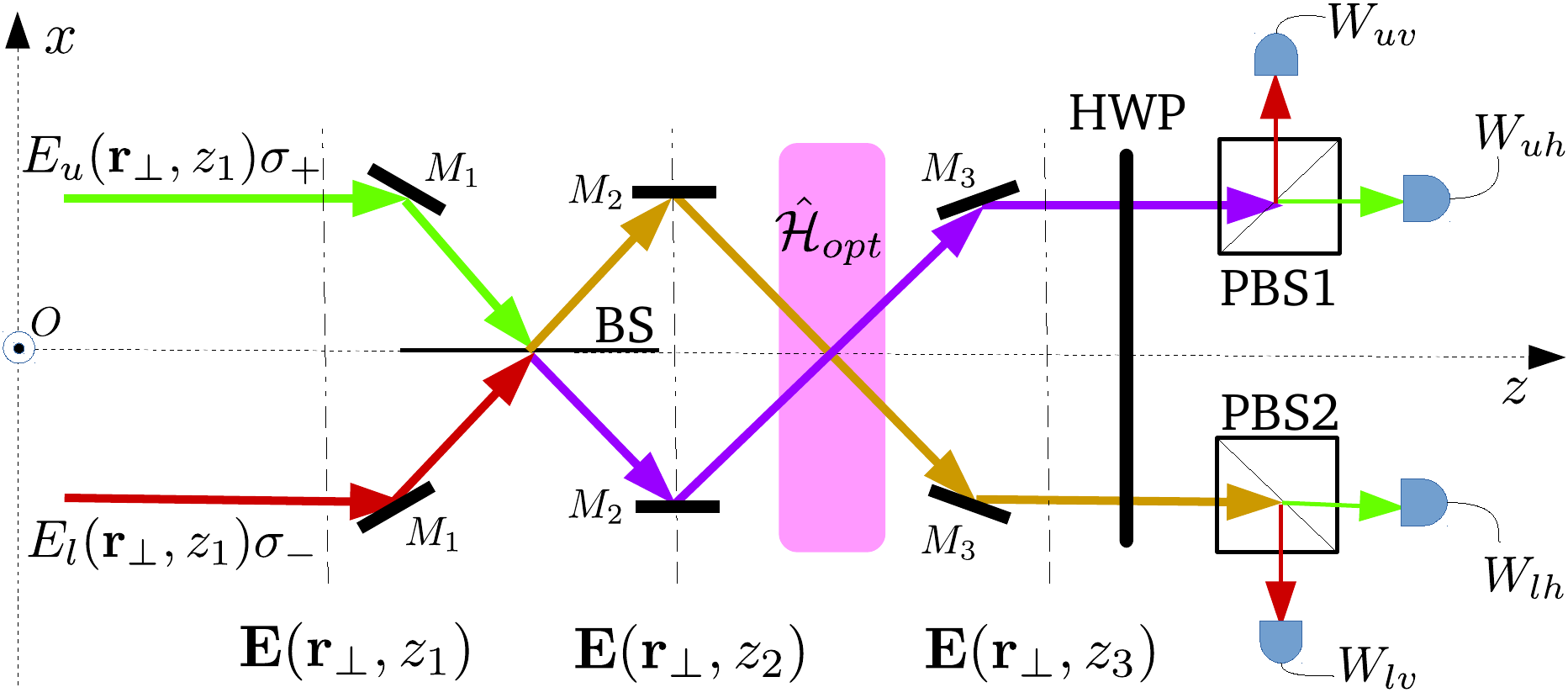}
\caption{(Color online) Schematic setup for testing the no-signaling principle with classical inseparable laser beams. The beam splitter BS and the half-wave plate HWP enable the choice of the local experimental measurement setting in the position and the polarization degree of freedom. $\hat{H}_{opt}$ is a PT-symmetric optical medium. PBS1 and PBS 2 are polarizing beams splitters which are part of the detection system. $M_i$ are mirrors to guide the laser beams.
}
\label{fig1}
\end{figure}

Here, we go one step further and show that the no-signaling principle can already be violated with classical inseparable laser beams in an optical local PT-symmetric system. This suggests that the problems associated to local $PT$-symmetric subsystems raised in Ref.~\cite{Lee2014PRL} are not exclusive to quantum mechanics, but already exist in the classical case. We also propose a possible implementation of the $PT$-symmetric subsystem using laser-controlled atoms.

In the quantum case, the two remote observers Alice and Bob act as independent  subsystems, which is possible due to the non-local nature of the entangled photon pair. Our classical setup illustrated in Fig.~\ref{fig1} does not allow for non-locality. Instead,  it considers polarization and position of laser beams as two independent subsystems $A$ and $B$, which are combined into a classical inseparable state. The beam splitter BS and the local non-Hermitian optical Hamiltonian $\hat{H}_{opt}$ act only on the position subsystem, while the half-wave plate HWP acts on the polarization subsystem. The polarization sensitive beam splitters PBS1 and PBS2 enable detection conditioned on position and polarization. $M_i$ are mirrors to guide the beams.  In the following, we discuss these components and our scheme in more detail.

{\it Two subsystems. }
As the first subsystem $A$, we choose the two-dimensional space spanned by the two mutually orthogonal circular polarization vectors $\bm{\sigma}_{+}=\textbf{e}_{h}+i\textbf{e}_{v}$ and $\bm{\sigma}_{-}=\textbf{e}_{h}-i\textbf{e}_{v}$. Here,  $\textbf{e}_{h}=\{1,0,0\}^T$ and $\textbf{e}_{v}=\{0,1,0\}^T$ are the corresponding horizontal and vertical polarization directions. To define the second subsystem $B$, we introduce two laser beams, which we label as upper ($u$) and lower ($l$) beam, respectively. The beams co-propagate in $z$ direction with equal frequencies, and are characterized by field amplitudes $E_u(\textbf{r}_{\perp},z)$ and $E_l(\textbf{r}_{\perp},z)$. We assume that the carriers of the field amplitudes do not overlap in space. Then, the polarization vectors and field amplitudes satisfy the orthogonality relations
\begin{subequations}
\label{orthogonality}
\begin{align}
 \bm{\sigma}^{*}_{m}\cdot\bm{\sigma}_{n}&=\delta_{mn}\,,\\
 \iint^{\infty}_{-\infty}E^{*}_{i}(\textbf{r}_{\perp},z_{1})E_{j}(\textbf{r}_{\perp},z_{1})d\textbf{r}_{\perp}&=\delta_{ij}\,, \label{orthogonality-field}
\end{align} 
\end{subequations}
where $\delta_{mn}~(m,n\in\{+,-\})$ and $\delta_{ij}~(i,j\in\{u,l\})$ are Kronecker delta functions. 
We can therefore interpret the polarization vectors $\{\bm{\sigma}_{+}, \bm{\sigma}_{-}\}$ and the field amplitudes $\{E_{u}(\textbf{r}_{\perp},z_{1}),  E_{l}(\textbf{r}_{\perp},z_{1})\}$ as orthonormal bases for two two-dimensional spaces. In the following, we will refer to them as polarization and position basis, respectively. 

For the subsequent analysis, we derive the representation of the parity operator in the position basis. To simplify the analysis, we choose $E_{u}(-\textbf{r}_{\perp},z)=E_{l}(\textbf{r}_{\perp},z)$. Then, 
\begin{subequations}
\label{parity-definition}
 \begin{align}
  &\hat{P}E_{u}(\textbf{r}_{\perp},z)= E_{u}(-\textbf{r}_{\perp},z) = E_{l}(\textbf{r}_{\perp},z)\,,\\[1ex]
  &\hat{P}E_{l}(\textbf{r}_{\perp},z)= E_{l}(-\textbf{r}_{\perp},z) = E_{u}(\textbf{r}_{\perp},z)\,,
 \end{align}
\end{subequations}
such that $\hat{P}$ simply swaps the two beams spanning the position space. In practice, the initial field amplitudes can be chosen, e.g., as $E_{u}(\textbf{r}_{\perp},z_{1}) = e^{-(\textbf{r}_{\perp}-\textbf{r}_{0})^2/w^2}$ and $E_{l}(\textbf{r}_{\perp},z_{1}) = e^{-(\textbf{r}_{\perp}+\textbf{r}_{0})^2/w^2}$ to satisfy Eq.~(\ref{parity-definition}). If further the beam displacement $\textbf{r}_0$ in the transverse plane is much larger than the beam waist $w$, i.e., $|\textbf{r}_{0}|\gg w$, then also Eq.~(\ref{orthogonality}b) can be satisfied.

{\it Initial state. }
As initial field, we choose 
\begin{align}
\label{Ein}
 \textbf{E}(\textbf{r}_{\perp},z_{1}) = E_{u}(\textbf{r}_{\perp},z_{1})\bm{\sigma}_{+} + E_{l}(\textbf{r}_{\perp},z_{1})\bm{\sigma}_{-}\,.
\end{align}
This input field is a classical  inseparable state, correlating the position and polarization subsystems. Such classical inseparable states are reminiscent of bipartite entangled states~\cite{Spreeuw1998FP,Lee2002PRL,Gabriel2011PRL,Qian2011OL,Kagalwala2013NPhoton,Aiello2015NJP,Berg-Johansen2015Optica,Song2015SciRep,Stoklasa2015NJP,Qian2015Optica}, but are not quantum entangled, since they lack nonlocality~\cite{Spreeuw1998FP,Karimi2015Science}. 
Classical inseparable beams have found applications in metrology~\cite{Toppel2014NJP,Berg-Johansen2015Optica} and information processing~\cite{Spreeuw2001PRA,Sun2015SciRep}, and also have proven valuable in visualizing abstract concepts in low-dimensional Hilbert spaces~\cite{Spreeuw1998FP,Fu2004PRA,Kagalwala2013NPhoton,Song2015SciRep,Qian2015Optica}.

{\it Experimental setting in position space. }
Using the mirror pair $M_1$, the input field is directed onto the BS, which acts on the position basis, but not on the polarization basis. It serves as the local experimental setting for the position subsystem. In the two-dimensional position space, its action can be represented by
\begin{align}
\label{BS}
 M_{BS}=&\begin{bmatrix}
          re^{i\theta_{1}} & te^{i\theta_{2}}\\[2ex]
          te^{i\theta_{3}} & re^{i\theta_{4}}
         \end{bmatrix}\,,
\end{align}
where $r$ and $t$ represent the reflection and transmission coefficients, respectively. Assuming a lossless BS, energy conservation imposes $r^2+t^2=1$ and $\theta_{2}-\theta_{1}+\theta_{3}-\theta_{4}=\pi$~\cite{Holbrow2002AJP}. In the following, we choose $\theta_{2}=\theta_{3}=0$ and $\theta_{1}=\theta_{4}=-\pi/2$.  
After the BS at the second mirror pair $M_{2}$, the total field can be written as
\begin{align*}
\label{EafterBS}
 \textbf{E}(\textbf{r}_{\perp},z_{2}) =& E_{u}(\textbf{r}_{\perp},z_{2})(-ir\bm{\sigma}_{+}+t\bm{\sigma}_{-}) \nonumber\\
                    &+E_{l}(\textbf{r}_{\perp},z_{2})(t\bm{\sigma}_{+}-ir\bm{\sigma}_{-})\,.
\end{align*}
Note that due to the beam splitter, the transmitted $E_{u}(\textbf{r}_{\perp},z_{1})$ merges with the reflected $E_{l}(\textbf{r}_{\perp},z_{1})$ to form $E_{l}(\textbf{r}_{\perp},z_{2})$ after the BS, and similarly for $E_{u}(\textbf{r}_{\perp},z_{2})$. 

{\it PT-symmetric medium. }
After the beam splitter, the field is directed to an optical medium, through which it propagates according to $i\frac{\partial}{\partial z} \textbf{E} = \hat{H}_{opt}\textbf{E}$. The optical Hamiltonian $\hat{H}_{opt}$ acts only on the position basis. Specifically, we assume a propagation dynamics as described by the coupled paraxial wave equations
\begin{subequations}
\label{wave-eqs}
 \begin{align}
 &i\frac{\partial E_{u}}{\partial z} = \left(-\frac{1}{2k}\frac{\partial^2}{\partial x^{2}} + \eta_{1}e^{i\phi_{1}}\right)E_{u} + \eta_{2}e^{i\phi_{2}} E_{l},\\[2ex]
 &i\frac{\partial E_{l}}{\partial z} = \left(-\frac{1}{2k}\frac{\partial^2}{\partial x^{2}} + \eta_{1}e^{-i\phi_{1}}\right)E_{l} + \eta_{2}e^{-i\phi_{2}} E_{u},
 \end{align}
\end{subequations}
where $k$ is wave numbers of the two beams. Real coefficients $\eta_{1},\eta_{2}$ and $\phi_{1},\phi_{2}\in[0,2\pi)$ are determined by the properties of the medium. The optical Hamiltonian $\hat{H}_{opt}$ then follows as 
 \begin{align}
\label{PT-Hamiltonian}
 \hat{H}_{opt}=&\begin{bmatrix}
	-\frac{1}{2k}\frac{\partial^2}{\partial \textbf{r}^{2}_{\perp}} + \eta_{1}e^{i\phi_{1}} & \eta_{2}e^{i\phi_{2}}\\[3ex]
        \eta_{2}e^{-i\phi_{2}}&  -\frac{1}{2k}\frac{\partial^2}{\partial \textbf{r}^{2}_{\perp}} + \eta_{1}e^{-i\phi_{1}}     
             \end{bmatrix}\,.
\end{align}
$\hat{H}_{opt}$ is Hermitian for $\phi_1 = 0$, but non-Hermitian otherwise. Since $\hat{H}_{opt}=\hat{P}\hat{T}\hat{H}_{opt}\hat{P}\hat{T}$, it is PT symmetric with $\hat{P}$ defined in Eq.~(\ref{parity-definition}) and $\hat{T}$ defined as $\hat{T}:~i\rightarrow-i$.

Note that the term $\partial^2/\partial \textbf{r}^{2}_{\perp}$ gives rise to paraxial diffraction, which may cause distortions to the spatial distribution of the two beams. However, in the final measurement, only the beam intensities are considered, and not their spatial shape. Thus, we neglect diffraction in the following analysis. Then, the optical propagation operator $M_{opt}(z)=e^{-i\hat{H}_{opt}z}$, evaluated at propagation distance $L$, becomes
\begin{align}
 M_{opt}\left(L\right)=~\frac
 {e^{-i\eta_{1}\cos(\phi_{1})L}}{\cos\alpha}\begin{bmatrix}
                                          \sin\alpha & -ie^{i\phi_{2}}\\[2ex]
                                          -ie^{-i\phi_{2}} & -\sin\alpha & \\
                                         \end{bmatrix}\,.
\end{align}
Here, $\sin\alpha=\eta_{1}\sin(\phi_{1})/\eta_2$. The medium length is given by $L=\pi/(2\eta_{2}\cos\alpha)$. We have further used $\eta_2>\eta_{1}|\sin\phi_{1}|$ such that $\alpha\in(-\pi/2,\pi/2)$.

After the optical medium, at mirror pair $M_3$, the field evaluates to
\begin{align}
 \textbf{E}(\textbf{r}_{\perp},z_{3}) =& \hat{\sigma}_{x}M_{opt}(L)\textbf{E}(\textbf{r}_{\perp},z_{3})\,,
\end{align}
where the Pauli operator $\hat{\sigma}_{x}$ acts on the position basis and reflects the fact that the upper and lower  beams have been interchanged while passing through the medium. Note that this exchange could be reversed with another set of mirrors, without changing the results of the subsequent analysis.

{\it Experimental setting in polarization space. }
As the last step before the measurement, the beam passes a half wave plate HWP which acts only on the polarization degree of freedom. The HWP determines the experimental setting in polarization space, and is characterized by a rotation in plarization space~\cite{Spreeuw1998FP} with $\textbf{e}_{h} \to \cos(\beta)\textbf{e}_{h}-\sin(\beta)\textbf{e}_{v}$ and $\textbf{e}_{v} \to \sin(\beta)\textbf{e}_{h}+\cos(\beta)\textbf{e}_{v}$.

{\it Detection. }
Finally, the beam is split into four output signals according to polarization ($h,v$) and position ($u,l$) via the polarizing beam splitters PBS1 and PBS2. At each of the four output ports, the respective field intensities are recorded. They can be evaluated to give 
\begin{subequations}
 \begin{align}
 &W_{lh}=\frac{1+\sin^{2}\alpha}{\cos^{2}\alpha}-rt\sin(2\beta)-I\,,\\[2ex]
 &W_{lv}=\frac{1+\sin^{2}\alpha}{\cos^{2}\alpha}+rt\sin(2\beta)+I\,,\\[2ex]
 &W_{uh}=\frac{1+\sin^{2}\alpha}{\cos^{2}\alpha}+rt\sin(2\beta)-I\,,\\[2ex]
 &W_{uv}=\frac{1+\sin^{2}\alpha}{\cos^{2}\alpha}-rt\sin(2\beta)+I\,,
\end{align}
\end{subequations}
where 
\begin{align}
 I = &\frac{\sin\alpha[r^2\sin(2\beta-\phi_{2})-t^2\sin(2\beta+\phi_{2})]}{\cos^{2}\alpha}\,.
\end{align} 
The $W_{nj}$ serve as joint measurements on the position ($n\in\{u,l\}$) and polarization ($j\in\{h,v\}$) degrees of freedom. Normalizing to the total intensity, we further define joint probabilities
\begin{align}
 P(n,j)=\frac{W_{nj}}{\sum_{n,j}W_{nj}}\,,
\end{align}
and the single probabilities of subsystem $A$ for polarization $j\in\{h,v\}$ become
 \begin{align}
 &P_A(j)=P(u,j)+P(l,j)
\end{align}
According to the no-signaling principle, the single probabilities for one subsystem  should not be affected by the choice of the measurement setting in the other subsystem. In particular, $P(h)$ and $P(v)$ should not depend on the properties of the beam splitter, i.e., should not be a function of $r$ and $t$. Nevertheless, we find that
\begin{subequations}
\label{violation}
\begin{align}
  P_A(h)=\frac{1}{2}-\frac{\sin\alpha[r^2\sin(2\beta-\phi_{2})-t^2\sin(2\beta+\phi_{2})]}{1+\sin^2\alpha}\,,\\[2ex]
 P_A(v)=\frac{1}{2}+\frac{\sin\alpha[r^2\sin(2\beta-\phi_{2})-t^2\sin(2\beta+\phi_{2})]}{1+\sin^2\alpha}\,.
\end{align}
\end{subequations} 
In general, $P_A(h)$ and $P_A(v)$ do depend on $r$ and $t$, such that the no-signaling principle is violated. The no-signaling principle only holds if $\alpha=0$, which corresponds to the case of a Hermitian optical Hamiltonian $\hat{H}_{opt}$. 

Note that the single probabilities related to the position degree of freedom $P_B(n)=P(n,h)+P(n,v)$ with $n\in\{u,l\}$ evaluate to $1/2$, such that they do not depend on the experimental setting $\beta$ in polarization space. This is reasonable because the non-Hermitian contribution only acts on position space. 

Finally, to test the classical nature of our system, we analyze a CHSH-like inequality~\cite{Clauser1969PRL,Aspect1982PRL,Tittel1998PRL,Weihs1998PRL,Hensen2015Nature}. Defining $C(\theta,\beta)= P(u,h)-P(u,v)-P(l,h)+P(l,v)$ and $S(\varphi_1,\beta_1,\varphi_2,\beta_2) = |C(\varphi_1,\beta_1)+C(\varphi_1,\beta_2) +C(\varphi_2,\beta_1)-C(\varphi_2,\beta_2)|$, we find
\begin{align}
S =\frac{\cos^2\alpha}{1+\sin^2\alpha}S_{0} \leq S_0\,.
\end{align}
Here,  $r=\sin\varphi$, and $S_{0}$ is the result in the Hermitian case $\alpha=0$. A direct calculation leads to $S_{0} \leq 2$, where $2$ is the classical bound. Since $S < S_{0}$, the CHSH-like quantity does not exceed the classical bound, even if the no-signaling principle is violated.

\begin{figure}[t!]
\includegraphics[width=8.0cm]{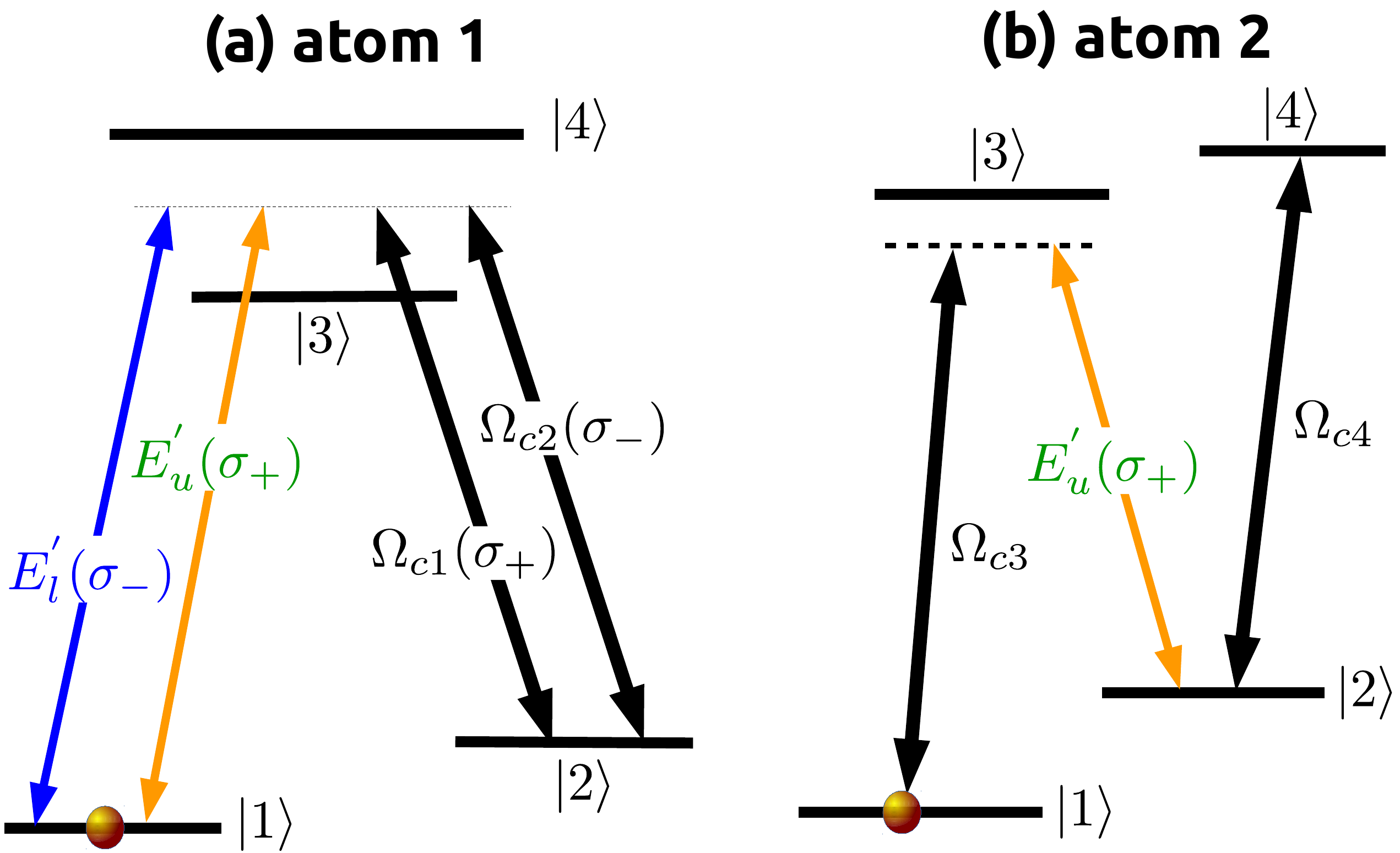}
\caption{(Color online) Realization of the optical PT-symmetric Hamiltonian based on a DFWM process. 
For the atomic species 1 in (a), the $F=2,m_F=0$ and $F=1,m_F=0$ sublevels of state $5 ^{2}S_{1/2}$ in  $^{87}\text{Rb}$ can serve as the two ground  states $|1\rangle$ and $|2\rangle$. The excited states $|3\rangle$ and $|4\rangle$ are chosen as $F=1, m_{F}=1$ and $F=2, m_{F}=-1$ sublevels of state $5 ^{2}P_{1/2}$.
The parameters are chosen as 
$\Gamma=2\pi\times 6~\text{MHz}$, 
$\Omega_{c1}=26.16\Gamma$, and
$\Omega_{c2}=10.0\Gamma$.
The detunings are 
$\Delta_{u}=\Delta_{c1}=-4.32\Gamma$ and
$\Delta_{l}=\Delta_{c2}=\Delta_{u}-\omega_{43}$ with $\omega_{43}=21.6\Gamma$ the frequency difference between $|3\rangle$ and $|4\rangle$. 
For species 2 in (b), $F=2, m_{F}=0$ and $F=3, m_{F}=0$ of $5 ^{2}S_{1/2}$ in $^{85}\text{Rb}$ are chosed for $|1\rangle$ and $|2\rangle$, and $F=3, m_{F}=1$ and $F=3~m_{F}=-1$ of $5 ^{2}P_{1/2}$ for $|3\rangle$, $|4\rangle$, respectively.
The parameters are 
$\Omega_{c3}=4.30\Gamma$,
$\Delta_{c3}=-22.2\Gamma$,
$\Omega_{c4}=0.03\Gamma$, and
$\Delta_{c4}=0$. 
With this configuration, one obtains $\eta_1 =1.91$, $\phi_1 = 0.84\pi$, $\eta_2 =36.5 $ and $\phi_2 =0 $ for atomic density $n_{0}=1.0\times 10^{11}~\text{cm}^{-3}$ and the abundance ratio $19.63:80.37$ between $^{87}\text{Rb}$ and $^{85}\text{Rb}$.}
\label{fig2}
\end{figure}

{\it Discussion. } We have demonstrated that the no-signaling principle can be violated in a classical system involving a non-Hermitian PT-symmetric subsystem. We therefore conclude that the problems associated to local $PT$-symmetric subsystems raised in Ref.~\cite{Lee2014PRL} cannot exclusively be linked to quantum mechanics, since they already exist in classical settings. This suggests that these problems are related to the peculiar properties of the $PT$-symmetric dynamics itself.

It should be noted that in our classical case, the beam polarization and its spatial profile used as the two subsystems are intrinsically inseparable. In particular, it is not possible to spatially separate them in order to perform independent measurements, which usually is a requirement for the violation of the no-signaling principle. From this a loophole arises in that a mechanism unaccounted for in our classical theory could lead to an exchange of information between the two subsystems without violating causality. However, we regard this loophole as the most unlikely option. 

A further analysis shows that the classical inseparable initial state and the PT-symmetric subsystem as such are not sufficient for the violation. For example, if the local PT-symmetric medium is placed in front of the BS, then the no-signaling principle is preserved. In this case, the PT-symmetric medium in essence only changes the initial state to a conventional optical setup with BS, HWP and the detection system. Since the latter optical system preserves the no-signaling theorem independent of the initial state, no violation is possible.

In order to experimentally test the violation of the no-signaling theorem in our classical setup, one has to implement the optical Hamiltonian Eq.~(\ref{PT-Hamiltonian}). This is possible by exploiting laser-matter interactions. A possible implementation is shown in Fig.~\ref{fig2}. Suppose that $E_{u}$ and $E_{l}$ propagate through an ensemble of atoms which consists of two different atomic species. Species 1 features a four-level double-lambda level structure. The two fields $E_{u}$ and $E_{l}$ couple to the transitions $|1\rangle \leftrightarrow |3\rangle$ and $|1\rangle \leftrightarrow |4\rangle$, respectively. Two additional resonant control fields with Rabi frequencies $\Omega_{c1}$ and $\Omega_{c2}$ enable a degenerate four-wave mixing~(DFWM) process. The control fiend parameters are chosen such that the conjugate anti-diagonal elements of $\hat{H}_{opt}$ are realized. $E_{u}$ is further coupled to species 2 with a $N$-type level scheme. Two additional control fields $\Omega_{c3}$ 
and $\Omega_{c4}$ create an active Raman gain (ARG) configuration such that the absorption experienced by $E_{u}$ from atomic species 1 is overcompensated. By suitably adjusting this gain, the complex conjugate diagonal elements of the optical Hamiltonian can be obtained.

This implementation crucially relies on the assumption that $E_u$ and $E_l$ can selectively be coupled to particular transitions. While it is possible in principle to exploit the different propagation directions of the two beams, a simpler approach is to use particular polarization states. This becomes possible, if one considers only two BS settings  (i) $r=1$ and $t=0$; (ii) $r=0$ and $t=1$, which are sufficient to observe the violation of no-signaling principle. In these two cases, $E_u$ and $E_l$ have definite and orthogonal polarization states, such that selective coupling becomes possible. It should be noted, however, that then the action of $H_{opt}$ depends on the polarization state. But it is possible to adjust the control field parameters in such a way that the same optical Hamiltonian is achieved for both cases (i) and (ii), see Fig.~\ref{fig2}.

In conclusion, we found that the no-signaling principle can be violated with classical inseparable laser beams in the presence of a local optical PT-symmetric subsystem. The possibility to implement local optical PT-symmetric subsystems via light-matter interactions enables the experimental exploration of problems associated to local PT symmetry, and provides a handle to subtle quantum concepts via classical analogues.

\bibliographystyle{apsrev4-1}
\bibliography{referencesbase}

\end{document}